\documentclass[aps,prd,onecolumn,amsmath,amssymb,nofootinbib,preprintnumbers]{revtex4}

\usepackage[pdftex]{graphicx}
\usepackage{epstopdf}
\usepackage{amsmath}
\usepackage{amssymb}
\usepackage{subfigure}
\usepackage{hyperref}
\usepackage{url}
\usepackage{xcolor}
\usepackage{color}
\usepackage{graphicx,color}
\usepackage{gensymb}
\usepackage{tabularx}

\begin{document}

\preprint{YITP-20-36}

\title{\textbf{A path(-integral) toward non-perturbative effects in Hawking radiation}}

\author{
Pisin Chen$^{a,b,c,d}$\footnote{{\tt pisinchen@phys.ntu.edu.tw}},
Misao Sasaki$^{a,e,f}$\footnote{{\tt misao.sasaki@ipmu.jp}}, 
and
Dong-han Yeom$^{g,h}$\footnote{{\tt innocent.yeom@gmail.com}}
}

\affiliation{
$^{a}$Leung Center for Cosmology and Particle Astrophysics, National Taiwan University, Taipei 10617, Taiwan\\
$^{b}$Department of Physics, National Taiwan University, Taipei 10617, Taiwan\\
$^{c}$Graduate Institute of Astrophysics, National Taiwan University, Taipei 10617, Taiwan\\
$^{d}$Kavli Institute for Particle Astrophysics and Cosmology,
SLAC National Accelerator Laboratory, Stanford University, Stanford, California 94305, USA\\
$^{e}$Kavli Institute for the Physics and Mathematics of the Universe (WPI), University of Tokyo, Chiba 277-8583, Japan\\
$^{f}$Yukawa Institute for Theoretical Physics, Kyoto University, Kyoto 606-8502, Japan\\
$^{g}$Department of Physics Education, Pusan National University, Busan 46241, Republic of Korea\\
$^{h}$Research Center for Dielectric and Advanced Matter Physics, Pusan National University, Busan 46241, Republic of Korea
}

\begin{abstract}
Hawking's seminal discovery of black hole evaporation was based on the semi-classical, perturbative method. Whether black hole evaporation may result in the loss of information remains undetermined. The solution to this paradox would most likely rely on the knowledge of the end-life of the evaporation, which evidently must be in the non-perturbative regime. Here we reinterpret the Hawking radiation as the tunneling of instantons, which is inherently non-perturbative. For definitiveness, we invoke the picture of shell-anti-shell pair production and show that it is equivalent to that of instanton tunneling. We find that such a shell pair production picture can help to elucidate firewalls and ER=EPR conjectures that attempt to solve the information paradox, and may be able to address the end-life issue toward an ultimate resolution.
\end{abstract}

\maketitle

%\newpage

There have been several approaches to justify the existence of Hawking radiation. The original Hawking derivation was based on the Bogoliubov transformation between the past infinity and the future infinity \cite{Hawking:1974sw}. Then the renormalized energy-momentum tensor was evaluated in order to better understand the physics near the horizon \cite{Davies:1976ei}. In addition to these two approaches, the particle tunneling picture was investigated \cite{Hartle:1976tp,Parikh:1999mf}. These descriptions are either consistent or equivalent. However, they are all based on the semi-classical, perturbative methods. One important consequence of black hole Hawking evaporation is the possible loss of information \cite{Hawking:1976sw}. This paradox has been under debate for more than 40 years without a clear resolution. One thing appears certain is that the final resolution relies on the knowledge of the black hole evaporation beyond the Page time \cite{Page:1993} and even toward its end-life (For a review, see, for example \cite{Chen:2015coy}). Since Hawking temperature is inversely proportional to the black hole mass, it is inevitable that there exists a crossing point beyond which the perturbative description becomes invalid and therefore a generalized description of Hawking radiation that encompasses both the perturbative and non-perturbative regimes is essential to provide us insights toward the final resolution of the information loss paradox. 

In principle, a non-perturbative description can be accomplished via the Euclidean path-integral approach \cite{Hartle:1983ai}. The transition amplitude from the in-state $(h^{\mathrm{in}}_{ab}, \phi^{\mathrm{in}})$ to the out-state $(h^{\mathrm{out}}_{ab}, \phi^{\mathrm{out}})$ can be described by the Euclidean path-integral \cite{Hartle:1983ai}
\begin{eqnarray}
\Psi \left[ h^{\mathrm{out}}_{ab}, \phi^{\mathrm{out}}; h^{\mathrm{in}}_{ab}, \phi^{\mathrm{in}} \right] = \int \mathcal{D}g_{\mu\nu} \mathcal{D}\phi \;\; e^{- S_{\mathrm{E}}[g_{\mu\nu},\phi]} \simeq \sum_{\mathrm{on-shell}} e^{- S_{\mathrm{E}}^{\mathrm{on-shell}}},
\end{eqnarray}
where all possible metric $g_{\mu\nu}$ and field $\phi$ configurations are summed over and the last expression ($\simeq$) is due to the steepest-descent approximation of summing over the on-shell solutions.

Let us consider the Einstein gravity with a free scalar field and its small fluctuations:
\begin{eqnarray}
S_{\mathrm{E}} = - \int dx^{4} \sqrt{g} \left[ \frac{1}{16\pi} \mathcal{R} - \frac{1}{2} \left( \nabla \phi \right)^{2} \right] + \int_{\partial \mathcal{M}} \frac{\mathcal{K} - \mathcal{K}_{o}}{8\pi} \sqrt{h} dx^{3},
\end{eqnarray}
where $\mathcal{R}$ is the Ricci scalar, $\mathcal{K}$ and $\mathcal{K}_{o}$ are the Gibbons-Hawking boundary term and that with the periodically identified Minkowski spacetime \cite{Gibbons:1976ue}, respectively. Note that this is a free scalar field without a potential. Since the Einstein equation for this system is simply $\mathcal{R} = 8\pi (\nabla \phi)^{2}$, the on-shell action of the volume integration should vanish. Therefore the boundary term is the only non-vanishing contribution to the on-shell Euclidean action. For an out-state, one can consider an out-going wave packet of the scalar field in the Lorentzian region. Then in general, a real-valued instanton that connects in- and out-states does not exist, and the scalar field becomes complex-valued. However, in order to satisfy the \textit{classicality} condition at the future infinity \cite{Hartle:2008ng}, one can impose the constraint that only the real part of the scalar field can reach the future infinity \cite{Chen:2018aij}. As a result, this complex-valued scalar field will be hidden outside the Euclidean region. After subtracting the boundary contribution at infinity and regularizing the cusp contribution at the horizon \cite{Gregory:2013hja}, one obtains the probability of the process \cite{Chen:2018aij}:
\begin{eqnarray}
P \simeq e^{4\pi (M - \omega)^{2} - 4\pi M^{2}}
\end{eqnarray}
where $M$ is the mass of the black hole and $\omega$ is the emitted energy. Assuming that the metric back-reaction is small, the condition $\omega \ll M$ must be satisfied. One then recovers the correct Boltzmann factor $P \sim e^{-8\pi M \omega}$ and Hawking temperature $T = 1 / 8\pi M$.

One very important lesson from this investigation is that the matter field can in general be complexified \cite{Hartle:2008ng,Alexanian:2008kd}. Thus the imaginary part of the matter field can effectively be identified as a negative energy flux \cite{Chen:2015ria} and be defined as the anti-matter field even for non-perturbative processes. Specifically, as a first step toward a fully non-perturbative treatment, we consider a non-perturbative shell, which is the thin-wall limit of a domain-wall structure of a scalar field, and its emission from a black hole, where the imaginary part of the complexified matter field can be treated as an \textit{anti-shell} with a negative tension.

\begin{figure}
\begin{center}
\includegraphics[scale=0.6]{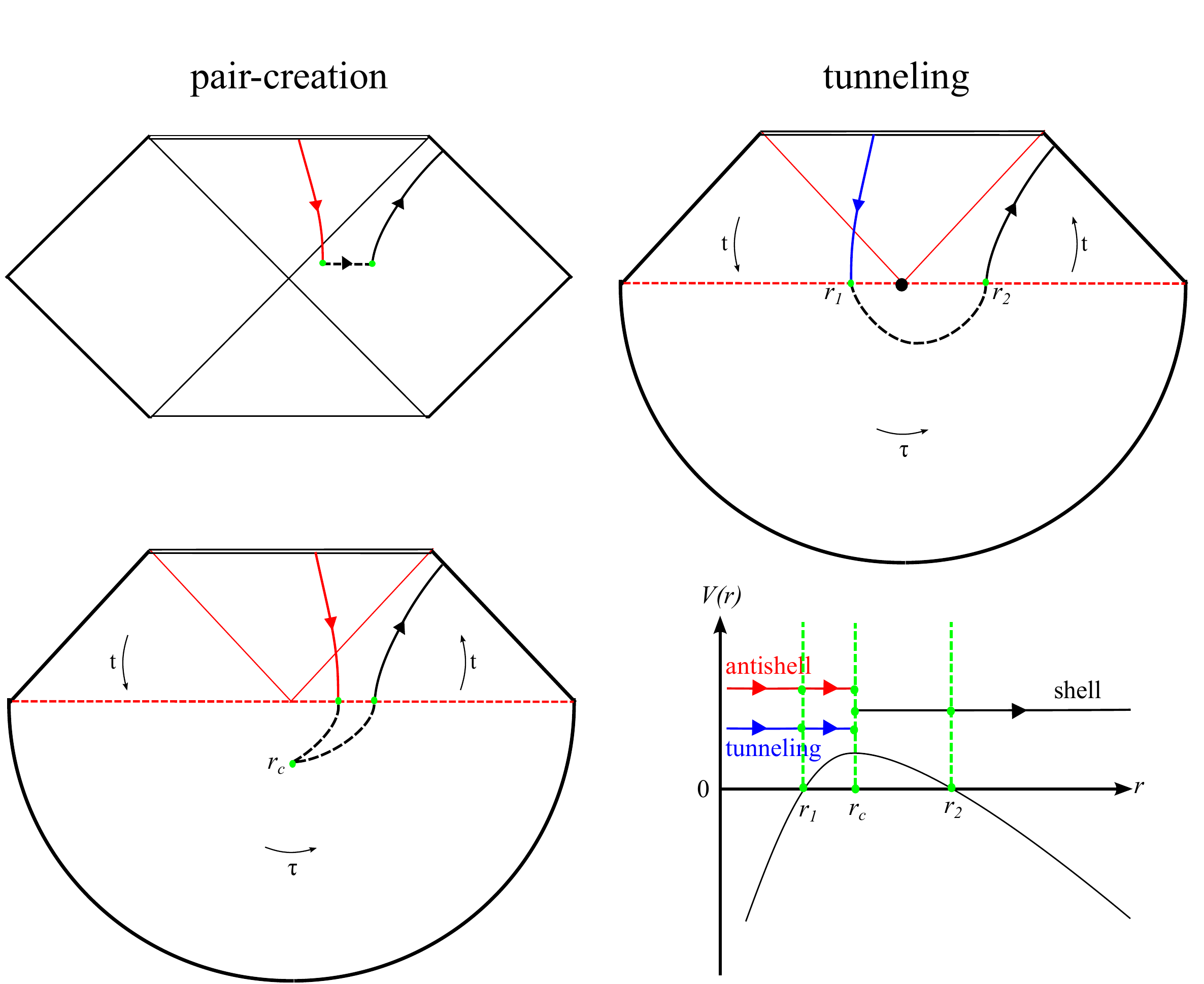}
\caption{\label{fig:newfig3}(Upper left) Shell-antishell pair-creation, where the two shells are created at different locations. (Upper right) The corresponding tunneling picture. (Lower left) This process can be interpreted as shell-anti-shell pair created at $r_{c}$ in the deep Euclidean regime. (Lower right) The effective potential, $V(r)$, with two turning points $r_{1}$ and $r_{2}$.}
\end{center}
\end{figure}

Keeping this in mind, let us consider the shell emission from a black hole. We consider the metric inside $(-)$ and outside $(+)$ the thin-shell as
\begin{eqnarray}
\label{eq:metric}
ds_{\pm}^{2}= - f_{\pm}(R) dT^{2} + \frac{1}{f_{\pm}(R)} dR^{2} + R^{2} d\Omega^{2},
\end{eqnarray}
while the thin-shell itself has the metric
\begin{eqnarray}
ds_{\mathrm{shell}}^{2} = - dt^{2} + r^{2}(t) d\Omega^{2}.
\end{eqnarray}
The Israel junction equation is then
\begin{eqnarray}\label{eq:junc}
\epsilon_{-} \sqrt{\dot{r}^{2}+f_{-}(r)} - \epsilon_{+} \sqrt{\dot{r}^{2}+f_{+}(r)} = 4\pi r \sigma,
\end{eqnarray}
where $\epsilon_{\pm} = +1$ if the area increases along the outward normal direction and $\sigma$ is the tension of the shell \cite{Israel:1966rt}. One can easily verify that if the junction equation is satisfied by a positive tension shell ($\sigma > 0$), then the same equation is also satisfied by a negative tension shell ($\sigma < 0$) under the flipping between the inside $(-)$ and the outside $(+)$. After simple calculations, the junction equation can be reduced to a simpler expression, $\dot{r}^{2} + V(r) = 0$ \cite{Blau:1986cw}, where the effective potential $V(r)$ is
\begin{eqnarray}\label{eq:form}
V(r) = f_{+}(r)- \frac{\left(f_{-}(r)-f_{+}(r)-16\pi^{2} \sigma^{2} r^{2}\right)^{2}}{64 \pi^{2} \sigma^{2} r^{2}}.
\end{eqnarray}
Note that the domain for $V(r) < 0$ follows the Lorentzian dynamics, while that for $V(r) > 0$ follows the Euclidean dynamics (lower right of Fig.~\ref{fig:newfig3}). In addition, if there exists a Euclidean domain in the system, then there must be two turning points (say, $r_{1} < r_{2}$).

Now the probability can be evaluated by the Euclidean action integral over the Euclidean manifold and the shell (upper right of Fig.~\ref{fig:newfig3}). In general, the action is composed of the following four parts \cite{Gregory:2013hja}:
\begin{eqnarray}
S_{\mathrm{E}} = (\mathrm{boundary\; term\; at\; infinity}) + (\mathrm{bulk\; integration}) + (\mathrm{contribution\; at\; horizon}) + (\mathrm{shell\; integration}),
\end{eqnarray}
where the contribution at horizon is due to the regularization of the cusp of the Euclidean manifold and the part for the shell integration has the form
\begin{eqnarray}\label{eq:shellint}
(\mathrm{shell\; integration}) = 2 \int_{r_{1}}^{r_{2}} dr r \left| \cos^{-1} \left( \frac{f_{+} + f_{-} - 16 \pi^{2} \sigma^{2} r^{2}}{2\sqrt{f_{+}f_{-}}} \right) \right|.
\end{eqnarray}
If one chooses the Euclidean time period as that where the shell bounces between the two turning points, then a cusp would emerge at the horizon. After performing the regularization \cite{Gregory:2013hja}, one obtains the contributions from the boundary term, the bulk term, and the horizon. The final result becomes
\begin{eqnarray}
\Delta S_{\mathrm{E}} = - \pi r_{f}^{2} + \pi r_{i}^{2} + (\mathrm{shell\; contribution}),
\end{eqnarray}
where $r_{i,f}$ correspond to the initial and final horizon radius, respectively.

This process can be equally appreciated from the shell-anti-shell pair production picture (upper left of Fig.~\ref{fig:newfig3}). Since in general the shell and the anti-shell are disconnected (lower left of Fig.~\ref{fig:newfig3}), the probability of this process should be given by the Euclidean solution that connects them. In this shell-anti-shell process over the Euclidean manifold, there must exist a point $r_{c}$ where the two shells collide. Since Eq.~(\ref{eq:shellint}) is invariant under the exchange of $\sigma \rightarrow - \sigma$ and $(\pm) \rightarrow (\mp)$, the anti-shell integration from $r_{1}$ to $r_{c}$ plus the shell integration from $r_{c}$ to $r_{2}$ is the same as that resulted from the shell tunneling picture. As the anti-shell with negative tension falls into the black hole, the black hole area will decrease, which results in the change of the action. On the other hand, the outgoing positive tension shell satisfies the classical junction equation, so it will not contribute to the probability of the process. One thus concludes that the probability for the infalling anti-shell to happen equals to the exponential with a negative exponent given by the difference between the initial and final black hole states, plus the Euclidean shell-anti-shell contribution \cite{Chen:2017suz}. It is important to remark that in this interpretation, we compare the action between that at the past infinity (initial state) and the future infinity (final state); therefore, regarding the final state, what we will evaluate is the black hole {\it after} the antishell collapsed. The shell-anti-shell and the tunneling interpretations (upper left and upper right of Fig.~\ref{fig:newfig3}) of Hawking radiation (in the non-perturbative regime) are therefore equivalent.

\begin{figure}
\begin{center}
\includegraphics[scale=1]{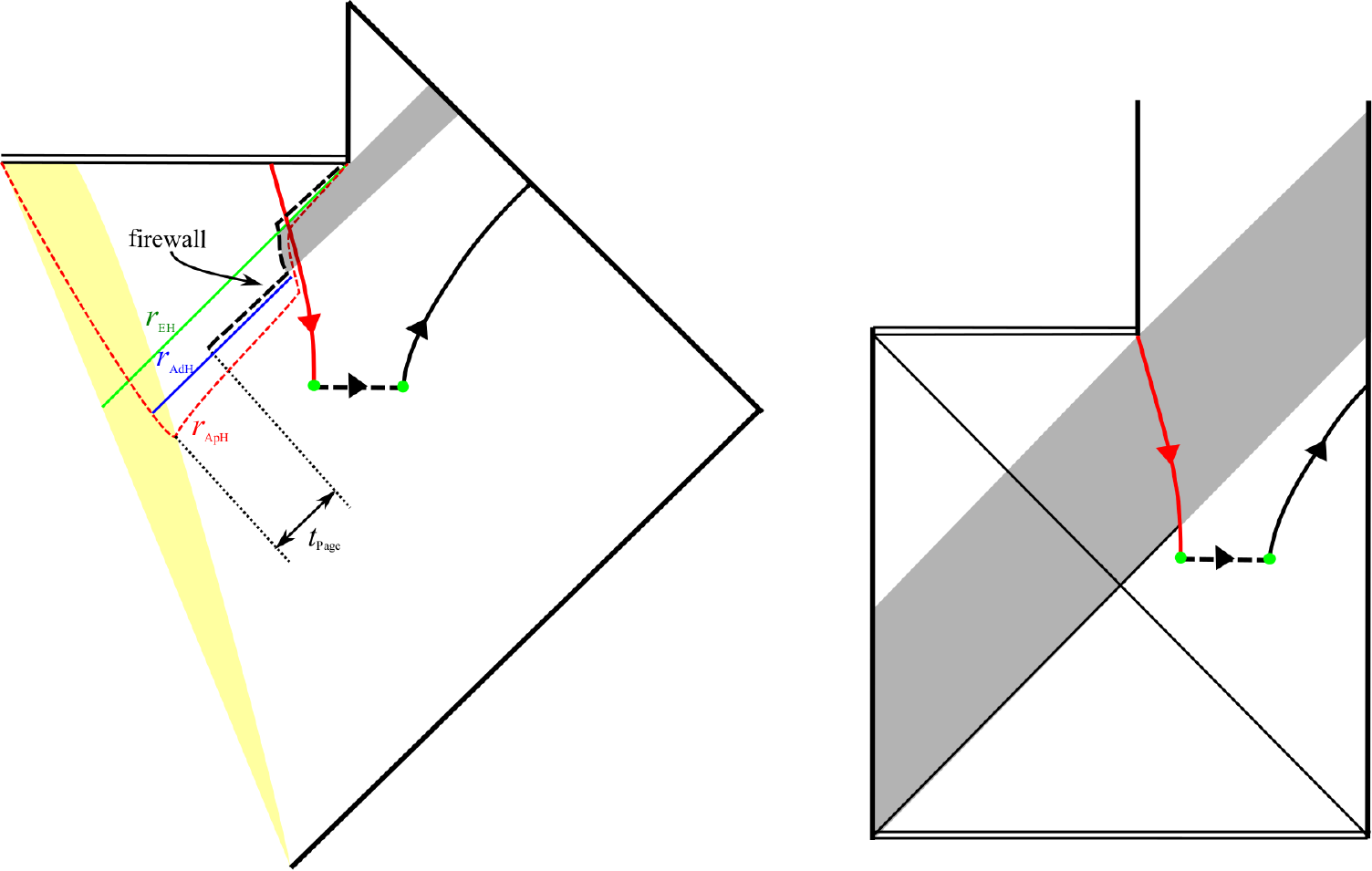}
\caption{\label{fig:newfig4}(Left) Criticism of the firewall proposal based on the shell-anti-shell pair creation picture of Hawking radiation, which argues that the firewall would become ``naked". (Right) Criticism of the ER=EPR conjecture based on the same picture, which argues that probability of traversing the Einstein-Rosen bridge is non-vanishing.}
\end{center}
\end{figure}

The notion of the negative tension shell makes easy the drawing of Penrose diagrams for various non-adiabatic quantum fluctuations \cite{Chen:2015gux} or non-perturbative processes \cite{Chen:2016nvj}. This would help clarify several conjectures about the information loss paradox such as the firewalls \cite{Almheiri:2012rt} and the ER=EPR  \cite{Maldacena:2013xja} conjectures. While the ``naked firewalls"  counter-argument \cite{Chen:2015gux} did not explicitly rely on non-perturbative effects, the notion of infalling negative tension anti-shells does help to make the argument of the detachment of the event horizon from the firewall more transparent (left of Fig.~\ref{fig:newfig4}), which indicates that firewalls would become visible to distant observers, and therefore it may not be as conservative as the original proponents claimed \cite{Almheiri:2012rt}. As for the ER=EPR conjecture \cite{Maldacena:2013xja}, under the shell-anti-shell pair creation picture (right of Fig.~\ref{fig:newfig4}), one can see more explicitly that the communication between the two sides of the Einstein-Rosen bridge is actually possible (gray colored region) albeit with a very small probability. This renders the ER=EPR conjecture questionable.

The Euclidean path-integral approach via the non-perturbative shell-anti-shell pair production interpretation of Hawking radiation may help to reach the ultimate resolution to the information loss paradox. For example, the exponentially scrambled but non-vanishing correlations may help to recover the original information \cite{Sasaki:2014spa}, following the intuitions of Hawking \cite{Hawking:2005kf}. When the mass of the black hole reduces to the Planck scale, the non-perturbative processes, which typically has the probability $\sim e^{-M^{2}/M_{\mathrm{Pl}}^{2}}$, would become comparable with that for the Hawking radiation. We speculate that this may provide a smooth transition of the geometry and the topology from a black hole to either a regular center, or an inert remnant at Planck size \cite{Adler:2001}, while in contrast the extrapolation of the perturbative description would predict a divergent and singular Hawking temperature toward the end-life of evaporation. 
%{\color{red}In real evaporating black holes, both of perturbative and non-perturbative effects will be superposed and the contribution from the latter part will be more and more important as the black hole mass decreases. Then, what does it look like if we sum-over all contributions? This is beyond the scope of this essay, but a very challenging and interesting question.}

%We conclude that the Euclidean path-integral approach via the non-perturbative shell-anti-shell pair production interpretation of Hawking radiation may lead us toward the ultimate resolution to the information loss paradox.

\section*{Acknowledgment}

This work was supported, for PC, by Ministry of Science and Technology (MOST) 107-2811-M-002-042, Taiwan; for MS, by the JSPS KAKENHI Nos. 15H05888 and 15K21733, and World Premier International Research Center Initiative (WPI Initiative), MEXT, Japan; and for DY, by the National Research Foundation of Korea (Grant No.: 2018R1D1A1B07049126).

\end{document}